\RequirePackage{fix-cm}
\documentclass[twocolumn]{svjour3}          % twocolumn
\smartqed  % flush right qed marks, e.g. at end of proof
\usepackage{graphicx}
\usepackage{subcaption}
\usepackage{amssymb}
\usepackage{amsmath}
\usepackage{multirow}
\usepackage{color}
\usepackage{algorithm}
\usepackage[noend]{algpseudocode}
\begin{document}

\title{Parallel Digital Predistortion Design on Mobile GPU and Embedded Multicore CPU for Mobile Transmitters}


\author{Kaipeng Li         \and
        Amanullah Ghazi \and
        Chance Tarver \and
        Jani Boutellier \and
        Mahmoud Abdelaziz \and
        Lauri Anttila \and
        Markku Juntti \and
        Mikko Valkama \and
        Joseph R. Cavallaro
}


\institute{Kaipeng Li, Chance Tarver and Joseph R. Cavallaro \at
              Department of Electrical and Computer Engineering\\
              Rice University, Houston, TX, 77005, USA\\
              \email{\{kl33,cat12,cavallar\}@rice.edu}           %  \\
           \and
           Amanullah Ghazi, Jani Boutellier and Markku Juntti\at
              Department of Computer Science and Engineering\\ University of Oulu, Finland
              \and
             Mahmoud Abdelaziz, Lauri Anttila and Mikko Valkama \at
             Department of Electronics and Communication Engineering\\
              Tampere University of Technology, Finland
}

\date{Received: date / Accepted: date}

\maketitle

\begin{abstract}
Digital predistortion (DPD) is a widely \\ adopted baseband processing technique in current radio transmitters. While DPD can effectively suppress unwanted spurious spectrum emissions stemming from imperfections of analog RF and baseband electronics, it also introduces extra processing complexity and poses challenges on efficient and flexible implementations, especially for mobile cellular transmitters, considering their limited computing power compared to basestations. In this paper, we present high data rate implementations of broadband DPD on modern embedded processors, such as mobile GPU and multicore CPU, by taking advantage of emerging parallel computing techniques for exploiting their computing resources. We further verify the suppression effect of DPD experimentally on real radio hardware platforms. Performance evaluation results of our DPD design demonstrate the high efficacy of modern general purpose mobile processors on accelerating DPD processing for a mobile transmitter.

\end{abstract}

\section{Introduction}
\label{intro}
During the development of low cost and efficient radio transceivers, direct conversion radio architecture, which relies on up-conversion and down-conversion of complex in-phase and quadrature (I/Q) signals, becomes popular in recent years \cite{txrx}, while entailing various impairments of the transceiver. Specifically, the imperfections of analog RF and digital baseband circuits of such transceivers can cause issues such as power amplifier (PA) nonlinearities, I/Q imbalance, local oscillator (LO) leakage, and so on. With the current trend on realizing massive multiple-input multiple-output systems \cite{mmimo}, impairments of each antenna may aggregate and even lead to severe signal distortion effects. 

At transmitter side, to achieve higher power efficiency and better signal coverage, the power amplifier (PA) is usually driven to its saturation region, where the nonlinearities are more pronounced, resulting in intermodulation distortion (IMD) and spurious spectrum emissions. I/Q imbalance and LO leakage can pose extra IMD terms at PA output. Therefore, problems such as violation of spurious emission limit for non-contiguous carrier aggregation (CA) transmission in 3GPP LTE-Advanced \cite{4g}, or violation of interference constraint between secondary user and primary user in cognitive radio systems \cite{cr} will arise if the spurious components are not well controlled. To solve above problems, people can simply resort to backing off the transmit power, which is called Maximum Power Reduction (MPR) \cite{mpr} in the context of 3GPP LTE uplink, to satisfy the limitation on spurious emission, while sacrificing the transmit efficiency and distance. 

Digital predistortion (DPD) technique is proposed and adopted recently as an alternative solution for spurious emission suppression by predistorting the I/Q samples at baseband before passing through the PA, so that harmful effects of transmitter impairments at the PA output can be mitigated \cite{dpd}. The parameters of a DPD algorithm will significantly affect the suppression results, and therefore, they should be well trained and estimated under certain transmitter circuit specifications and transmission environments. To deal with distortion effects of PA nonlinearities, I/Q imbalance and LO leakage all together, joint PA and I/Q modulator calibration and DPD parameter estimation are shown to be a promising approach for achieving effective DPD suppression without extra RF hardware \cite{joint,joint2}. 

Applying DPD on practical transmitters demands efficient and flexible implementations to meet the data rate requirement of modern wireless communication standards and to adapt to various transmit scenarios, such as single LTE carrier or non-contiguous LTE component carriers (CC). Although DPD is now a de-facto solution for modern basestations in cellular radio networks, the design and implementation of DPD on mobile transmitters have not been fully explored. With the increasing computing power of modern embedded processors, DPD designs for mobile transmitter can be feasible and expected to deliver high performance.

Over the past decade, mobile computing techniques and mobile applications have evolved rapidly thanks to the enhanced computing capability and portability of mobile system-on-chip (SoC) \cite{mobile}. Modern mobile SoC chipset usually integrates various embedded processors, such as multicore CPU, mobile GPU, and other application specific coprocessors. Specifically, general purpose computing on multicore CPU and GPU \cite{gpu} has become a new trend for accelerating signal processing and data analysis applications, such as computer vision \cite{cv}, machine learning \cite{ml} and wireless communication \cite{basestation,gpumimo}, with the development of parallel programming tools and models, such as pthreads and OpenMP for multicore CPU, or CUDA \cite{cuda} and OpenCL \cite{opencl} for mobile GPU, which provide higher facilitation and flexibility on exploiting the parallel architecture and numerous computing resources than conventional hardware such as FPGAs, DSPs and ASICs.

There are some previous work of DPD implementations on FPGA~\cite{fpga_dpd}, and transport trigger architecture (TTA)~\cite{tta}, which is an application-specific architecture like ASIC. Although those implementations can achieve good data rate performance and energy efficiency, they lack the design flexibility to easily reconfigure adaptive DPD parameters. The efforts to apply new parameters, recompile and resynthesis those designs are not trivial. In contrast, a DPD design on general purpose processor, such as CPU or GPU, can achieve comparable data rate performance if their computing resources are fully exploited, and provide high design flexibility and portability using high-level parallel programming languages, mature coding tools and short recompilation and rebuilding time. However, few DPD implementations have been done on such general purpose processors, especially mobile processors considering our DPD design targets mobile transmitters. Our previous work~\cite{mobilegpu}, to the best of the authors' knowledge, is the first CUDA-based DPD implementation on GPU, and this paper extends from our previous mobile GPU based DPD implementation with further design optimization and thus higher data rate, and details another embedded CPU-based design for comparison. \cite{opencl_dpd} also proposes an alternative implementation on mobile GPU using OpenCL.

\paragraph{Contributions:} In this paper, we are motivated to develop high performance DPD implementations targeting mobile transmitters by exploring the emerging parallel programming techniques and computing capability of modern parallel mobile processors. Specifically, on an ARM multicore CPU, we implement the DPD design using NEON single-instruction multiple-data (SIMD) intrinsics \cite{neon}, while on a mobile GPU, we provide an alternative DPD implementation based on CUDA. Furthermore, we integrate our DPD implementations on a novel software-defined mobile transmitter platform built by Jetson embedded development board \cite{tk1,tx1} and WARP v3 radio board \cite{warp}, and experimentally test the DPD suppression effect with real transmitter radio hardware. We benchmark the data rate performance of our two DPD implementations and monitor the PA outputs with a spectrum analyzer. Our results show the feasibility and efficiency for driving DPD on mobile transmitters by modern embedded processors, generate useful benchmark results on profiling  mobile GPU and CPU, and serve as case studies for future mobile signal processing applications in the context of wireless communication and Internet of things.

The paper is organized as follows. Section 2 overviews the DPD algorithm for implementation. Section 3 describes the implementation details and optimization strategies of DPD on both mobile GPU and multicore CPU. Section 4 demonstrates the DPD functionality experimentally on a real mobile transmitter platform. We show the performance evaluation results in Section 5 and conclude in Section 6.

\begin{figure*}[t]
\centering
\begin{subfigure}{.35\textwidth}
  %\centering
  \includegraphics[width=\linewidth]{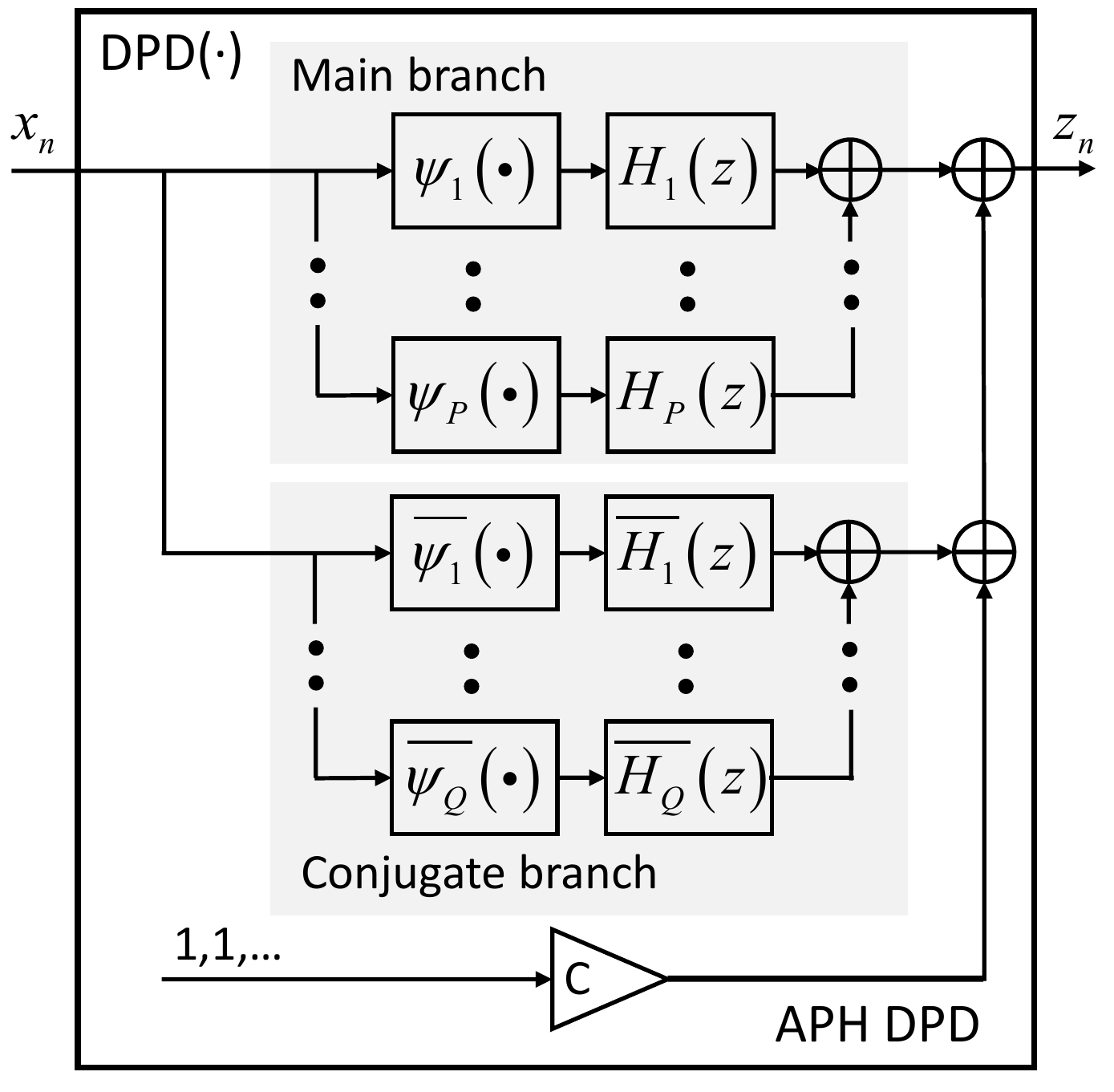}
  \caption{APH DPD structure}
\end{subfigure}%
\begin{subfigure}{.35\textwidth}
  %\centering
  \includegraphics[width=\linewidth]{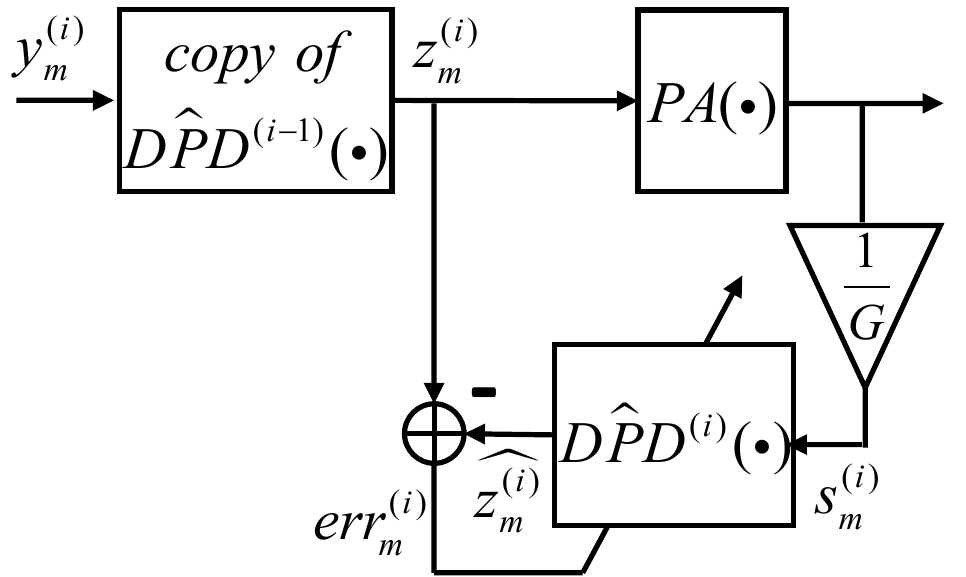}
  \caption{Indirect learning architecture}
\end{subfigure}
\caption{DPD architecture}
\end{figure*}

\section{Overview of DPD Algorithms}
We focus on a broadband DPD algorithm which performs joint mitigation of power amplifier and I/Q modulator impairments \cite{joint}, which is shown to deliver effective suppression effect on unwanted spectrum emissions according to simulation results. This algorithm includes two stages: a parameter estimation stage to generate DPD filter coefficients based on iterative training, and a predistortion stage for applying finalized DPD on actual streaming transmit samples. Compared to typical digital predistorters which require concatenated step-by-step processing for PA predistortion, LO leakage compensation, and I/Q mismatch predistortion with corresponding parameter estimation separately, the key idea of this joint mitigation based DPD approach is that it integrates the filtering operations in both PA predistorter and I/Q mismatch predistorter into an equivalent combined filtering operation, and regards the LO leakage compensation as an additional filter coefficient, so that the parameter estimation, i.e, the filter coefficient estimation, of DPD can be performed jointly as an one-shot estimation. 

\subsection{DPD Processing Structure}
Figure 1(a) shows the internal structure of the predistorter, the so-called augmented parallel Hammerstein (APH) structure. Consider that we have $N$ modulated I/Q samples $x_0, x_1, \cdots, x_{N-1}$ to transmit. Instead of directly passing them through the PA and radio hardware to the air, we can first send them as the input to the well-trained predistorter for necessary DPD processing, so that the predistorted samples $z_0, z_1, \cdots, z_{N-1}$ at the DPD output can be finally transmitted with compensation of transmitter impairments. The input-output relationship of the predistorter on a certain sample $z_n$ and $x_n$ $(n=0,1,\cdots, N-1)$ can be formulated as:

\begin{align}
\begin{split}
z_n&=DPD(x_n)\\
&=\sum_{\mbox{\tiny$p \in I_P$}}\sum_{\mbox{\tiny$k=0$}}^{L_p} h_{p,k}\psi_p(x_{n-k})\\
&+\sum_{\mbox{\tiny$q \in I_Q$}} \sum_{\mbox{\tiny$k=0$}}^{L_q} \bar{h}_{q,k}\bar{\psi}_q(x_{n-k})+c
\end{split}
\end{align}

According to the DPD structure described in Figure 1(a) and Equation (1), the APH DPD processing contains three major steps: polynomial computation denoted as $\psi(\cdot)$ indicating nonlinearity, filtering computation denoted as $H(\cdot)$, and accumulation of filtering results and LO leakage compensation $c$.

Specifically, $\psi_p(x_n)=\sum_{m \in I_p}{u_{m,p}|x_n|^{p-1}x_n}$ $(p \in I_P)$ and $\bar{\psi}_q(x_n)=\psi_q(x_n^*)=\sum_{m \in I_q}{u_{m,q}|x_n|^{q-1}x_n^*}$ $(q \in I_Q)$ are the polynomial of direct signal $x_n$ for $p^{th}$ main branch and the polynomial of conjugate signal $x_n^*$ for $q^{th}$ conjugate branch, respectively. $I_P$ and $I_Q$ denote the set of used polynomial orders in main and conjugate branch, respectively. For example, if only odd order polynomials are considered, then $I_P=\{1,3,5,\cdots,P\}$ and $I_Q=\{1,3,5,\cdots,Q\}$, where $P$ and $Q$ indicate the highest polynomial orders in the main branch and conjugate branch, respectively. Typically, $P$ is larger than $Q$ considering the conjugate signal caused by I/Q imbalance is usually weaker than direct signal, for example, $P=5$ and $Q=3$. $I_p$ and $I_q$ are the subset of $I_P$ and $I_Q$, containing polynomial orders up to $p$ and $q$, respectively, and $u_{m,p}$ and $u_{m,q}$ are corresponding coefficients of statistically orthogonal polynomials, which are pre-calculated according to \cite{poly}. $h_{p,k}$ and $\bar h_{q,k}$ denote the $k^{th}$ filter coefficient of FIR filter $H_{p}(z)$ with $L_p$ taps and FIR filter $\overline{H}_q(z)$ with $L_q$ taps, respectively. Those filter coefficients as well as LO leakage compensation $c$ are DPD parameters to be estimated during the iterative training stage before they are finally used for effective predistortion processing on actual payload samples.

\subsection{DPD Parameter Estimation}
For DPD parameter estimation, we pack those filter coefficients into vectors $\textbf{h}_\textbf{p}$=$[h_{p,0}\,h_{p,1}\cdots h_{p,L_p-1}]^T$ and $\bar{\textbf{h}}_\textbf{q}$=$[\bar{h}_{q,0}\, \bar{h}_{q,1}\cdots \bar{h}_{q,L_q-1}]^T$, and then stack  $\textbf{h}_\textbf{p}$ and $\textbf{h}_\textbf{q}$ for all $p$ and $q$ as well as parameter $c$ as a single coefficient vector \textbf{h}=[$\textbf{h}_\textbf{1}^\textbf{T}$, $\textbf{h}_\textbf{3}^\textbf{T}$ $\cdots$ $\textbf{h}_\textbf{P}^\textbf{T}$ $\bar{\textbf{h}}_\textbf{1}^\textbf{T}$ $\bar{\textbf{h}}_\textbf{3}^\textbf{T}$ $\cdots$ $\bar{\textbf{h}}_\textbf{Q}^\textbf{T}$ $c$]$^T$. In fact, the parameter estimation of DPD is to calculate the vector $\textbf{h}$ which can lead to optimal or near optimal DPD suppression effect at PA output for transmit samples, and can be realized by indirect learning architecture (ILA) \cite{ila}. As shown in Figure 1(b), ILA performs iterative training in a feedback loop. For a certain $i^{th}$ iteration, $M$ training samples $y_0^{(i)}, y_1^{(i)}, y_2^{(i)} \cdots y_{M-1}^{(i)}$ are prepared as the input of the DPD function $\hat{DPD}^{(i-1)}(\cdot)$, which is estimated from the ($i$-1)$^{th}$ iteration, and the DPD output samples $z_0^{(i)}, z_1^{(i)}, z_2^{(i)} \cdots z_{M-1}^{(i)}$, where $z_m^{(i)}=\hat{DPD}^{(i-1)}(y_m^{(i)})$ , are sent to the PA ($z_m^{(1)}=y_m^{(1)}$ in the first iteration). In the feedback loop, we extract the PA output samples $s_0^{(i)}, s_1^{(i)}, s_2^{(i)} \cdots s_{M-1}^{(i)}$ scaled by PA gain $G$ and estimate the parameters for $\hat{DPD}^{(i)}$ by least squares (LS) estimation. Specifically, the LS estimation gives $\hat{\textbf{h}}$ $^{\mathbf{(i)}}$=$(\mathbf{\Psi^H\Psi)}^{-1}\mathbf{\Psi^Hz^{(i)}}$ to minimize the sum of squared errors between current reference DPD output $\mathbf{z^{(i)}}$ in the TX path and to-be-estimated DPD output $\mathbf{\hat{z}^{(i)}}$  in the feedback path. Here, $\mathbf{\Psi^H}$ is the conjugate transpose of $\mathbf{\Psi}$; $\mathbf{z^{(i)}}$=$\mathbf{\Psi}\hat{\textbf{h}}^{\mathbf{(i-1)}}$, where $\hat{\textbf{h}}^{\mathbf{(i-1)}}$ indicates the estimated \textbf{h} from ($i$-1)$^{th}$ iteration and $\mathbf{\Psi}$ is the basis matrix defined as $\mathbf{\Psi}$=$ [\mathbf{\Psi}_1 \mathbf{\Psi}_3 \cdots \mathbf{\Psi}_P \overline{\mathbf{\Psi}}_1 \overline{\mathbf{\Psi}}_3 \cdots \overline{\mathbf{\Psi}}_Q\, \mathbf{1}]^T$. Element $\overline{\mathbf{\Psi}}_q$=$\mathbf{\Psi}_q^*$, and $\mathbf{\Psi}_p$ (or $\mathbf{\Psi}_q^*$) is the element matrix, defined as: 

\begin{align}
\begin{split}
\scriptsize
\mathbf{\Psi}_p=\begin{pmatrix} 
\psi_p(y_0)&0&0&\cdots&0\\
\psi_p(y_1)&\psi_p(y_0)&0&\cdots&0\\
\psi_p(y_2)&\psi_p(y_1)&\psi_p(y_0)&\cdots&0\\
\cdots&\cdots&\cdots&\cdots&0\\
\psi_p(y_{M-1})&\psi_p(y_{M-2})&\psi_p(y_{M-3})&\cdots&\psi_p(y_{M-L_{p}})\\
0&\psi_p(y_{M-1})&\psi_p(y_{M-2})&\cdots&\psi_p(y_{M-L_{p}}+1)\\
0&0&\psi_p(y_{M-1})&\cdots&\psi_p(y_{M-L_{p}}+2)\\
\vdots&\vdots&\vdots& &\vdots\\
0&0&0&\cdots&\psi_p(y_{M-1})\\
  \end{pmatrix}.  
\end{split}
\end{align}

For $(i+1)^{th}$ iteration, we can insert the estimated $\hat{DPD}^{(i)}(\cdot)$ in the TX path and perform similar estimation, and usually 1-3 iterations are enough for the convergence of the final parameter $\mathbf{h}$, which is to be used in the actual predistortion stage. 

With fixed PA and radio hardware, and relatively static environment, the DPD parameters can be trained and estimated offline and used for a long period without the need of retraining, while the actual predistortion with finalized filter coefficients, as structured in Figure 1(a), demands high data rate for streaming transmit samples. Therefore, in the following section, we focus on implementation details of the finalized digital predistorter on mobile processors, assuming that the training process has been completed offline and the final DPD parameters are ready to use for the implementations. Once a retraining is performed, we can simply reconfigure the values of DPD parameters in design functions and rebuild the design. We note, however, if retraining is frequently needed, for example, in an unusually fluctuating environment, we can also implement the training part on mobile processors to realize online training. We can use similar parallel programming techniques detailed in the following, while introducing extra computation complexity and processing latency for training, which we leave for future work to build a more complete software-defined mobile terminal.

\section{DPD Implementation on Parallel Mobile Processors}

In this section, we detail the DPD implementation on mobile processors targeting mobile transmitters. To map the DPD algorithm on parallel processors efficiently, we need to first explore the inherent data parallelism and data dependencies in the DPD algorithm. We then utilize a particular vectorization scheme for a specific processor, for example, GPU or multicore CPU, to realize the parallel data computation, and perform necessary but low-overhead communication to handle data dependencies. In Algorithm 1, we summarize the DPD processing algorithm to be implemented. We show specifications of our experimental mobile processors in Section 3.1, discuss the data parallelism in the DPD algorithm and vectorization schemes on mobile processors in Section 3.2, and describe how we handle the data dependencies and data communications efficiently by various optimization strategies in Section 3.3.

\begin{algorithm}[t]

\caption{DPD Processing Algorithm}
\begin{algorithmic}[1]
\small

\State \textbf{Input}: \\
{\quad}$x_n, n=0,1,...,N-1$; $c$;\\ 
{\quad}$h_{p,k}, p\in I_P, k=0,1,...,L_p-1$;\\ 
{\quad}$h_{q,k}, p\in I_Q, k=0,1,...,L_q-1$;

\State \textbf{Polynomial Computation}: \\ 
{\quad}$\psi_p(x_n)=\sum_{m \in I_p}{u_{m,p}|x_n|^{p-1}x_n}$ $(p \in I_P)$\\
{\quad}$\bar{\psi}_q(x_n)=\psi_q(x_n^*)=\sum_{m \in I_q}{u_{m,q}|x_n|^{q-1}x_n^*}$ $(q \in I_Q)$\\
{\quad}($u_{m,p},u_{m,q}$ are pre-calculated poly. coefficients)

\State \textbf{Filtering Computation}: \\ 
{\quad}$f_p(x_n)=\sum_{\mbox{\tiny$k=0$}}^{L_p} h_{p,k}\psi_p(x_{n-k})$\\
{\quad}$\bar{f}_q(x_n)=\sum_{\mbox{\tiny$k=0$}}^{L_q} \bar{h}_{q,k}\bar{\psi}_q(x_{n-k})$

\State \textbf{Accumulation Computation}: \\ 
{\quad}$z_n=\sum_{\mbox{\tiny$p \in I_P$}}f_p(x_n)+\sum_{\mbox{\tiny$q \in I_Q$}}{\bar{f}_q(x_n)}+c$

\State \textbf{Output}: \\
{\quad}$z_n, n=0,1,...,N-1$

\end{algorithmic}
\end{algorithm}

\subsection{Experimental Embedded Platform}
We implement the predistorter on mobile GPU using CUDA, and on embedded ARM multicore CPU based on NEON SIMD intrinsics with OpenMP multi-threading.  We benchmark our GPU and CPU implementations on two generation of Nvidia Jetson development boards, i.e, Jetson TK1 and TX1, for performance comparison. The specifications of the implementation platforms are listed in Table 1.

\begin{table}[t]
\centering

\caption{Specifications of the implementation platforms}
\scriptsize

\begin{tabular}{c|c|c}
\hline 
 & \textbf{Jetson TK1} & \textbf{Jetson TX1}\tabularnewline
\hline 
\hline 
SoC & 28nm Tegra K1 & 20nm Tegra X1\tabularnewline
\hline 
\multirow{2}{*}{CPU} & quad-core Cortex-A15  & quad-core Cortex-A57\tabularnewline
\cline{2-3} 
 & 32-bit ARMv7  & 64-bit ARMv8 \tabularnewline
\hline 
GPU & 192-core Kepler GPU & 256-core Maxwell GPU\tabularnewline
\hline 
Coding & \multicolumn{2}{c}{CUDA for GPU / NEON+OpenMP for CPU}\tabularnewline
\hline 
Compiler & \multicolumn{2}{c}{nvcc -O3 for GPU / GCC -O3 for CPU}\tabularnewline
\hline 
OS & \multicolumn{2}{c}{Linux for Tegra (L4T)}\tabularnewline
\hline 
\end{tabular}

\end{table}

\subsection{Data Parallelism Exploration}

\begin{figure}[t]
\centering
  \includegraphics[width=0.45\textwidth]{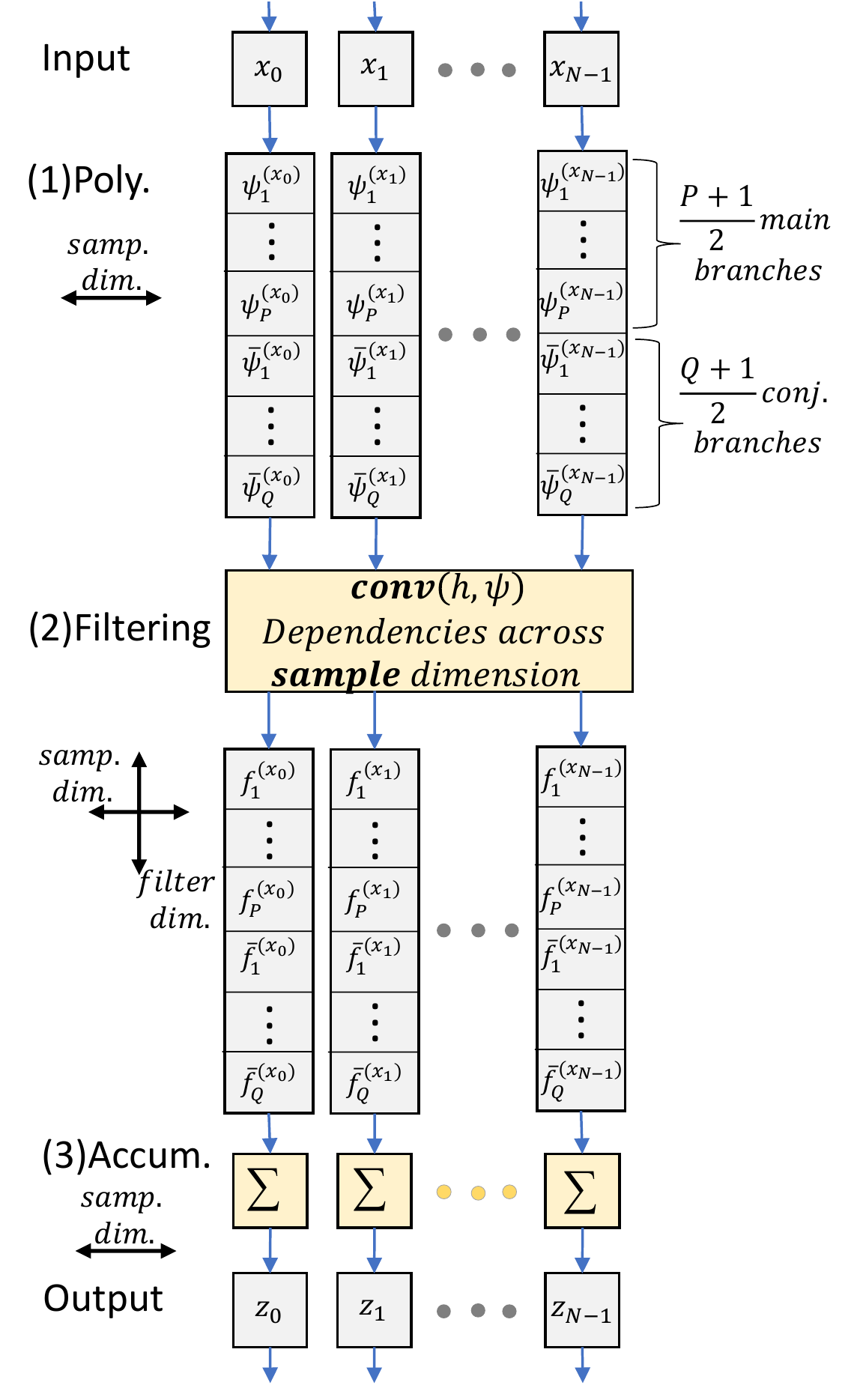}
\caption{Data flow and parallelism}       % Give a unique label
\end{figure}

\subsubsection{Parallelism Analysis}
According to Figure 1(a) and Algorithm 1, to obtain a certain predistorted output sample $z_n$ from a certain input sample $x_n$, we have three major steps: (1)polynomial computation: calculate the polynomial results at each of $P$ main branches and $Q$ conjugate branches for a window of $L_p$ or $L_q$ input samples, respectively; (2)filtering computation: calculate the filtering result for each branch (filter) based on the estimated filter coefficient and the corresponding window of polynomial results; (3)accumulation: accumulate the filtering result from each branch as well as the LO leakage compensation $c$ to generate the final output $z_n$. Assume we have $N$ input samples for DPD processing. In step (1), for a certain input sample $x_n$, the polynomial computation for higher order is dependent on lower order polynomial results. Although we can still calculate the polynomials for each order independently on a certain $x_n$, a more efficient way to avoid redundant computations is to calculate the polynomial from low order to high order in serial for both main branch and conjugate branch, so that the only data parallelism in this step is that we can calculate polynomials for each sample $x_n$ in parallel with total parallelism degree of $N$ across sample dimension. In step (2), since we already have all the polynomial results, when we calculate the filtering result for a certain branch $p$ or $q$, we can extract the corresponding window of polynomial results from step (1), for example, $\psi_p(x_{n-L_p+1})$, $\psi_p(x_{n-L_p+2})$, $\cdots$, $\psi_p(x_{n-1})$, $\psi_p(x_n)$ for branch $p$ corresponding to input sample $x_n$. Therefore, for filtering computation, we introduce another parallel dimension and have even higher data parallelism: we can calculate the filtering result for each branch (filter) across filter dimension and for each sample across sample dimension in parallel with total parallelism degree $N\times R$, where $R=(P+1)/2+(Q+1)/2$ indicates the total number of branches (filters) in DPD. In step (3), all the filtering results for a certain sample $x_n$ need to be reduced to one accumulation in serial, but we can calculate the accumulation for each sample $x_n$ in parallel with parallelism degree of $N$ across sample dimension. Figure 2 visualizes the data flow and parallelism for each step discussed above.

\subsubsection{Data Vectorization on GPU}
On the GPU, we implement the computing flow by CUDA kernel functions, and invoke the kernel with large number of parallel threads to perform the computation in parallel based on single instruction multiple threads (SIMT) execution model. In our previous work \cite{mobilegpu}, we have designed three kernels to perform polynomial computation, filtering computation and accumulation, which are invoked with $N$, $NR$, $N$ threads, respectively, indicating their parallelism degrees. However, in this way, we need to share intermediate results between those kernels using GPU global memory, which will pose significant memory access overhead. Alternatively, we can combine those kernels into a single kernel invoked with $N$ threads to perform the DPD processing under a per-sample basis, while wasting some parallelism degree for the filtering computation. In fact, when we have a large number of $N$ samples to process with $N$ threads, the GPU performance will saturate with high occupancy of cores, a larger $NR$ parallelism will benefit little. However, a combined kernel can effectively reduce the memory access overhead by sharing the intermediate results via faster shared memory or even local registers, which is shown to achieve a performance gain for the whole DPD design. 

\subsubsection{Data Vectorization on multicore CPU}
On the multicore CPU, we can realize the data parallelism by two-level vectorization: (1)thread level: since there are four high-power cores on the ARM CPU of Jetson TK1 and Jetson TX1, we can generate four threads using OpenMP, each controlling the DPD processing for one fourth of the total workload running on each core; (2)instruction level: NEON SIMD instructions, an advanced SIMD extension in ARM processors, operating on 64-bit doubleword or 128-bit quadword NEON vector registers, are supported in the ARMv7 and ARMv8 architecture. For using NEON SIMD instructions, we can either code low level NEON assembly instructions with manually controlled instruction selection and scheduling, or code high-level NEON intrinsics which serve as function calls of C/C++ programs and leave the instruction selection and scheduling to the compiler. Here, in our design, we choose NEON intrinsics considering the high facilitation on design development, and high portability for different ARM CPUs, rather than using NEON assembly, which needs to be specifically optimized for a certain ARM CPU. The input and output of a NEON intrinsic function usually require a special vector data type. We utilize the 128-bit quadword registers in the register bank on NEON unit to represent a vector of four 32-bit floating point elements, for example, we define the real part or imaginary part of complex input samples using the \texttt{float32x4\_t} data type of NEON, so that when a compiled NEON instruction operates on a certain \texttt{float32x4\_t} data, it actually processes four floating point elements in parallel with the single instruction. For the DPD processing, most of the computation operations are additions and multiplications, which can be easily realized by \texttt{vaddq\_f32} and \texttt{vmulq\_f32} intrinsics, where \texttt{q} indicates 128-bit quadword and \texttt{f32} indicates 32-bit floating point elements in a vector. Based on such NEON SIMD intrinsics, every four 32-bit floating point data can be processed together in parallel under a per-sample basis, on each core controlled by an OpenMP thread, and therefore we have 16 samples in total to process at the same time on the multicore CPU.

For a certain input sample, while the data dependencies within polynomial computation of different orders and data dependencies for accumulation of different branches can be handled simply and efficiently in serial, the data dependencies within filtering computation are more tricky, since the polynomial results for a window of $L_p$ or $L_q$ samples are required, but not only the polynomials of the current sample. Therefore, when we process the input samples under a per-sample basis in parallel, for example, for input sample $x_n$, we need to explore an efficient way to prepare and extract the necessary window of polynomial results for obtaining the filtering results indexed by $x_n$. We discuss our strategies for handling such filtering dependencies in the following section.

\begin{figure*}[t]
\centering
  \includegraphics[width=0.8\textwidth]{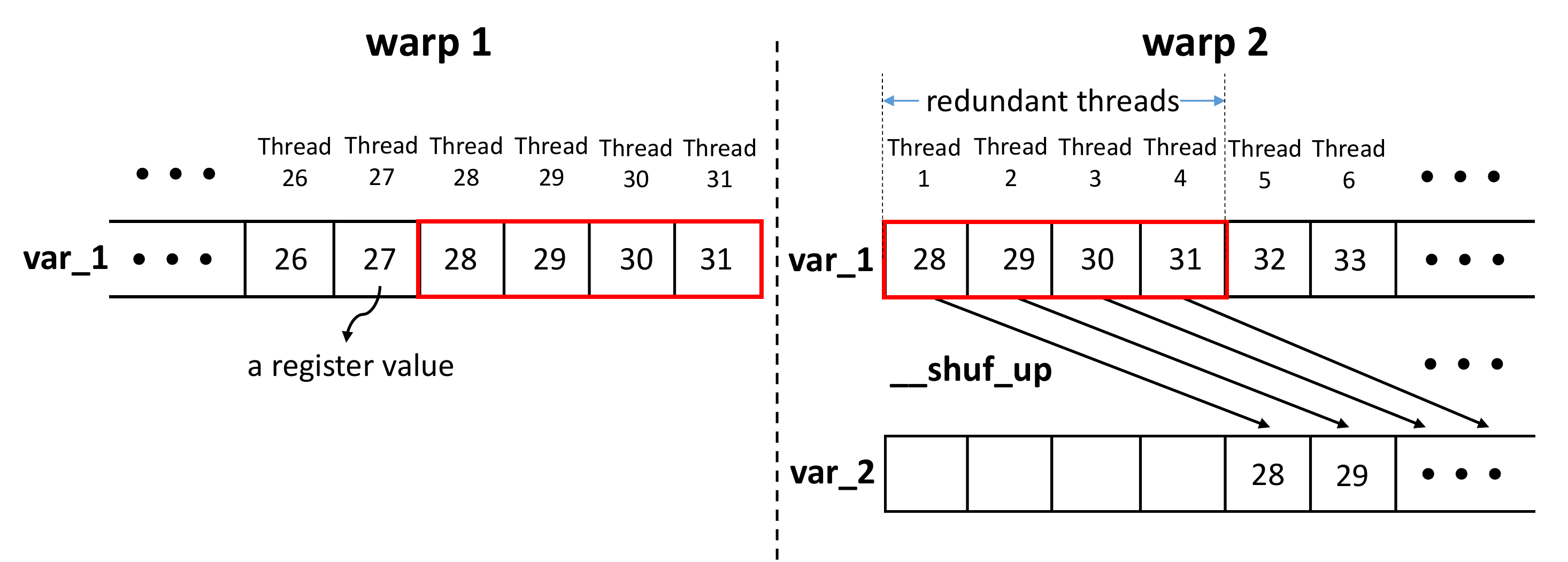}
\caption{Warp shuffle}       % Give a unique label
\end{figure*}

\subsection{Memory Access Optimization}
To enhance the performance of the DPD design, efficient memory access needs to be taken care of for necessary data sharing and communication. On Tegra SoC, CPU and GPU share a unified device memory, and have their own local caches and registers. The key goal for memory access optimization is to resort to the slow device memory only when necessary, but to exploit the faster on-chip local caches and near-core registers for communication as much as possible while carefully ensuring that we keep their utilization under resource limitations considering their scarcity.

\subsubsection{Zero Copy Access of Device Memory on GPU}
On the GPU, since we combine all the DPD processing computations into a single kernel as discussed before, we only need to fetch the data from device memory as DPD input, and store back the processed data to device memory as DPD output, and other intermediate results should be buffered and shared in local caches or registers. For DPD input and output data, since they should essentially reside in CPU memory, a conventional way for GPU to access them is to perform explicit memory copy from CPU memory to GPU memory for the input and from GPU memory to CPU memory for the output. Such explicit memory copy will lead to significant overhead and thus degradation of data rate performance. Usually, on desktop GPU, one can schedule and pipeline the GPU kernel execution and CPU-GPU memory copy in multiple streams, so that the CPU-GPU memory copy latency can be overlapped and the kernels in multiple streams can also execute concurrently to improve performance. Unfortunately, the Jetson TK1 or TX1 board, which has only one streaming multiprocessor on chip and a unified memory, currently does not support multi-stream scheduling. Actually, considering the CPU and GPU share a unified on-board physical memory, we can use another strategy, \emph{zero copy access} \cite{cuda}, to avoid the explicit CPU-GPU memory copy by mapping the CPU host memory address to GPU device memory pointer, which can be then passed to the kernel function. To enable zero copy access, we should begin with setting device flag by \texttt{cudaSetDeviceFlags(cudaDeviceMapHost)}, and allocate host memory using \texttt{cudaHostAlloc} function call with special flag \texttt{cudaHostAllocMapped}, and then map the host memory to GPU device pointer by \\ \texttt{cudaHostGetDevicePointer} function call. In this way, the kernel will directly fetch the input from the host memory which the mapped GPU device pointers point to and store back the output to the host memory similarly without explicit host-device memory copy. Those extra configurations and function calls for zero copy access can be set up at the beginning of the program before performing any computations, thus pose little extra overhead on the data rate performance of actual DPD processing.

For multicore CPU, the memory access is straightforward: within a certain core controlled by an OpenMP thread, we can load a  vector containing four 32-bit floating point elements from memory to 128-bit NEON registers via \texttt{vld1q\_f32} SIMD intrinsic and store the output back from NEON registers to memory via \texttt{vst1q\_f32} SIMD intrinsic, where flag \texttt{1q} indicates one 128-bit quadword vector.

\subsubsection{Inter-thread Communication via Warp Shuffle on GPU} 
As discussed in Section 3.2, when we perform DPD computation for each input sample $x_n$ in parallel, we should be aware of the data dependencies during the filtering computation: the FIR filter at $p^{th}$ main branch or $q^{th}$ conjugate branch requires a window of $L_p$ or $L_q$ polynomial results of both current and previous input samples. A simple and direct way is to use GPU device memory which can be accessed by any invoked thread in any thread block to buffer the intermediate results, that is, when we complete the polynomial computations for all samples, we store the polynomial results back to GPU device memory, and for a certain FIR filter, for example, the $p^{th}$ main branch filter when processing $x_n$ in thread $n$, which requires the polynomial results $\psi_p(x_{n-L_p+1})$, $\psi_p(x_{n-L_p+2})$, $\cdots$, $\psi_p(x_{n-1})$, $\psi_p(x_n)$, we extract them again from global memory within that thread $n$ for the following filtering. However, this approach will lead to extra memory access overhead and competitions. A better way is to use shared memory, a special L1 cache which can be controlled by the programmer, to store the intermediate polynomial results which can be accessed by threads within the same thread block, but it is still far slower than local registers and can also arise issues such that several adjacent threads compete for a shared polynomial result.

Here, in our design, to achieve optimized data sharing and communication between threads, we utilize \emph{warp shuffle} technique, which was introduced in Kepler GPU ~\cite{shuffle}, to realize direct register-to-register data shuffling among different threads within a thread $warp$. Specifically, we call \texttt{dest\_var=\_\_shuf\_up(source\_var, delta, warpSize)} in a thread to retrieve a certain register variable \texttt{source\_var} from the thread whose index is smaller than the calling thread by a number of \texttt{delta}, so that the retrieved \texttt{dest\_var} in the calling thread can be used for the following computations. \texttt{warpSize} is a fixed number of 32 reserved by Nvidia. In our problem, during the filtering computation for a certain sample $x_n$, we can use such \texttt{\_\_shuf\_up} intrinsic to retrieve the polynomial results corresponding to previous $L_p-1$ or $L_q-1$ input samples calculated by lower-indexed near-neighbor threads, for the thread which controls the DPD processing for current $x_n$. Since that \texttt{\_\_shuf\_up} can only be used for thread communication within a $warp$, the first $L_p-1$ or $L_q-1$ threads in a certain $warp$ cannot access all of their required $L_p$ or $L_q$ polynomial results which may reside in another $warp$. To resolve this, we can simply overlap some computations between two neighbor $warps$ with consecutive thread index numbers. For example, if we set $L_p=5, L_q=5, \forall p, q$, then the first $warp$, which includes $32$ threads, will perform the DPD processing for sample $x_0$ to $x_{31}$; for the second $warp$, it will operate on input samples $x_{28}, x_{29}, x_{30}, x_{31}, x_{32},\cdots, x_{59}$, where the processing of $x_{28}$ to $x_{31}$ by the first $4$ threads in the second $warp$ is redundant and only for obtaining the polynomial results required by the \texttt{\_\_shuf\_up} called from the $5^{th}$ to $8^{th}$ threads in that $warp$. Similarly, the third $warp$ will operate on sample $x_{56}$ to $x_{87}$ with its first $4$ threads as redundant threads. When $L_p$ and $L_q$ are small, such as $5$, the performance gain obtained from direct register-to-register data shuffle for resolving dependencies without accessing shared or global memory is shown to be more significant than some small overhead introduced by such redundant threads. Figure 3 shows our \emph{warp shuffle} approach for achieving efficient inter-thread communication.

For the estimated filter coefficients, since they are pre-calculated before DPD processing, we can simply fetch and store them in the shared memory, so that all threads within a thread block can access them efficiently for the filtering computation.

\begin{figure}[t]
\centering
  \includegraphics[width=0.48\textwidth]{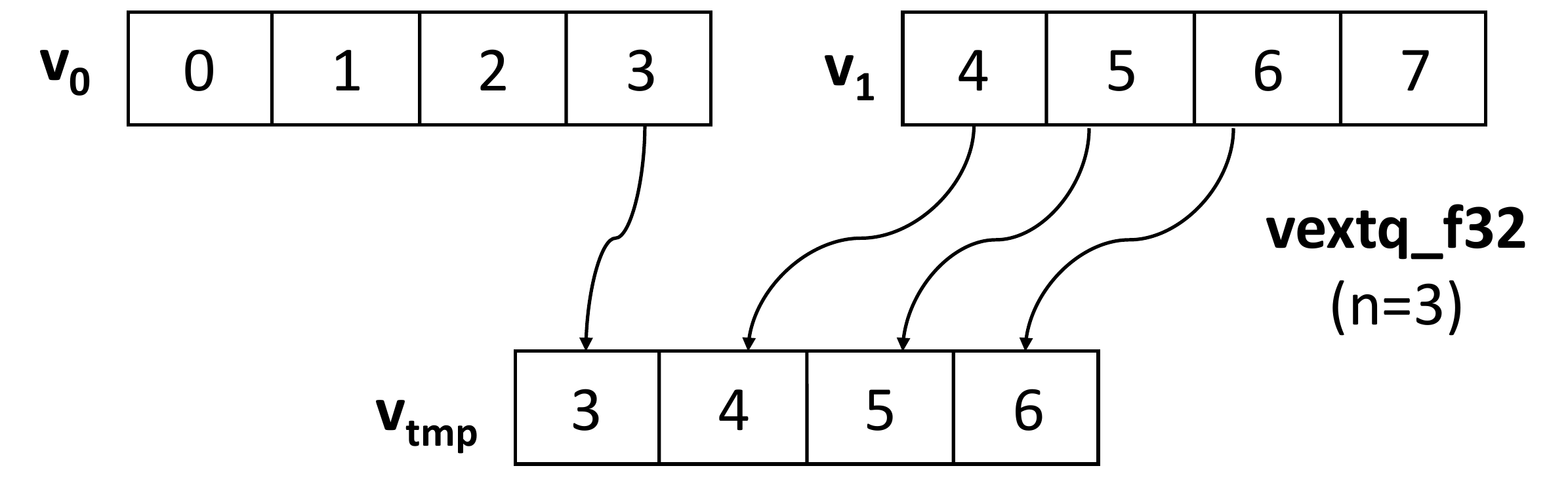}
\caption{Vector extraction}       % Give a unique label
\end{figure}

\subsubsection{Inter-vector Data Regrouping via Vector Extraction on CPU} 
Similar to the GPU implementation, for the CPU implementation, to avoid unnecessary global memory access, we only load the input data from memory before DPD computation and store back the DPD output to memory after the computation, but exploit faster NEON registers and general local registers for buffering and sharing intermediate results during the computation.

For the filtering computation which exposes data dependencies on polynomial results of previous samples, we may simply store the polynomial results of all input samples back to CPU memory after polynomial computation, and extract the required window of results for the following filtering, while sacrificing the performance from significant global memory access overhead. In fact, when we have $N$ DPD input samples, instead of generating $N$ threads to process them all together like GPU, we have $N/4$ input samples deployed by OpenMP on each of the four cores, and process them vector by vector, each including 4 samples, based on SIMD instructions, so we need $N/16$ iterations in serial for completing the processing of all $N/4$ samples on each CPU core. Therefore, in our design, we buffer polynomial results of one or two previous iterations, which correspond to polynomial results of previous samples, to NEON registers as temporary variables, and combine with the polynomial results in the current iteration to extract the necessary ones for filtering computation. For example, when $L_p=5$ and $L_q=5$, the first iteration computes the polynomial results at each branch (we omit the branch index $p,q$ for simplicity) for first four input samples as a vector $\mathbf{v_0}$=\{$\psi(x_0)$, $\psi(x_1)$, $\psi(x_2)$, $\psi(x_3)$\} by SIMD instructions, and the second iteration computes $\mathbf{v_1}$=\{$\psi(x_4)$, $\psi(x_5)$, $\psi(x_6)$, $\psi(x_7)$\}. For processing input samples $x_4, x_5, x_6, x_7$ in this iteration, at the first filtering tap, we need to prepare $\mathbf{v_{tmp}}$=\{$\psi(x_4)$, $\psi(x_5)$, $\psi(x_6)$, $\psi(x_7)$\} interacting with the first filter coefficient $\mathbf{f_1}$=\{$h_1$, $h_1$, $h_1$, $h_1$\}, each element corresponding to each input, and at the second filtering tap, we need to have $\mathbf{v_{tmp}}$=\{$\psi(x_3)$, $\psi(x_4)$, $\psi(x_5)$, $\psi(x_6)$\} interacting with the second filter coefficient $\mathbf{f_2}$=\{$h_2$, $h_2$, $h_2$, $h_2$\}, and so on. Here, vector $\mathbf{v_{tmp}}$ in the second iteration can be extracted from the buffered $\mathbf{v_0}$ and current $\mathbf{v_1}$ by using  $\mathbf{v_{tmp}}=$\texttt{vextq\_f32}($\mathbf{v_0,v_1},n$) NEON intrinsic to pack the lower-end $n$ elements of $\mathbf{v_1}$ and the $4-n$ higher-end elements of $\mathbf{v_0}$ into $\mathbf{v_{tmp}}$, within NEON registers, as shown in Figure 4. For larger $L_p$ and $L_q$, we can buffer the polynomial results from more previous iterations and perform the similar vector extraction to regroup inter-vector data for filtering computation.

\subsection{Design Summary and Comparison}
In this part, we summarize and compare the major design decisions of DPD implementations on mobile GPU and multicore CPU.

In Table 2, we show a typical configuration of APH DPD parameters which can achieve good DPD suppression effect. We use this configuration for the experimental verification and performance benchmark in the following sections. Considering the reconfigurability of our implementations, we can easily update those parameters, for example, with higher order polynomials and more filter branches, for possible better DPD suppression effect on certain radio hardware and in certain transmit environment.

\begin{table}[t]
\centering{}%
\caption{APH DPD configuration\label{tab:config}}
\scriptsize
\begin{tabular}{c|c|c}
\hline 
\multirow{1}{*}{\textbf{Parameter}} & \textbf{Main branch} & \textbf{Conjugate branch}\tabularnewline
\hline 
\hline 
Max polynomial order & $P$=5  & $Q$=3\tabularnewline
\hline 
Number of filters & 3 & 2\tabularnewline
\hline 
Taps per filter & $L_p$=5 (for each $p$) & $L_q$=5 (for each $q$)\tabularnewline
\hline 
\end{tabular}
\end{table}

In Table 3, we summarize and compare the major techniques and optimization strategies for enhancing the performance of our GPU and CPU implementations. While using different programming models and schemes, they share the same goal to exploit the data parallelism and to facilitate the data access and communication for better performance.

\begin{table}
\centering{}%
\caption{Comparison of implementation techniques}
\scriptsize
\begin{tabular}{c|c|c}
\hline 
 & \textbf{Mobile GPU} & \textbf{Multicore ARM CPU}\tabularnewline
\hline 
\hline 
Exec. model & SIMT & SIMD\tabularnewline
\hline 
Data type & 32-bit FP & 128-bit quadword vector\tabularnewline
\hline 
Vectorization  & CUDA threads & OpenMP+NEON intrinsics\tabularnewline
\hline 
Parallelism  & N threads & 4 threads $\times$ 4 samples/vector\tabularnewline
\hline 
Memory access & zero copy & \texttt{vld1q\_f32/vst1q\_f32} \tabularnewline
\hline 
Data Sharing  & warp shuffle & \texttt{vextq\_f32}\tabularnewline
\hline 
\end{tabular}
\end{table}

Our current embedded DPD designs are targeting mobile transmitters, however, we emphasize that our implementations are portable and scalable to desktop GPUs and CPUs if we want to apply DPD at basestations. On desktop GPUs which support CUDA, we can invoke even more threads and thread-blocks to realize data parallelism on thousands of CUDA cores, and can perform multi-stream scheduling for pipelining CPU-GPU memory copy and kernel execution as an alternative technique of zero copy to resolve memory copy overhead. Our previous work~\cite{mobilegpu} has discussed DPD performance on desktop GPUs as reference. On desktop CPUs, we can take advantage of even longer SIMD instructions, such as SSE and AVX, with 256-bit or 512-bit registers, and generate more OpenMP threads on more CPU cores for higher performance.

\begin{figure}[t]
\centering
  \includegraphics[width=0.48\textwidth]{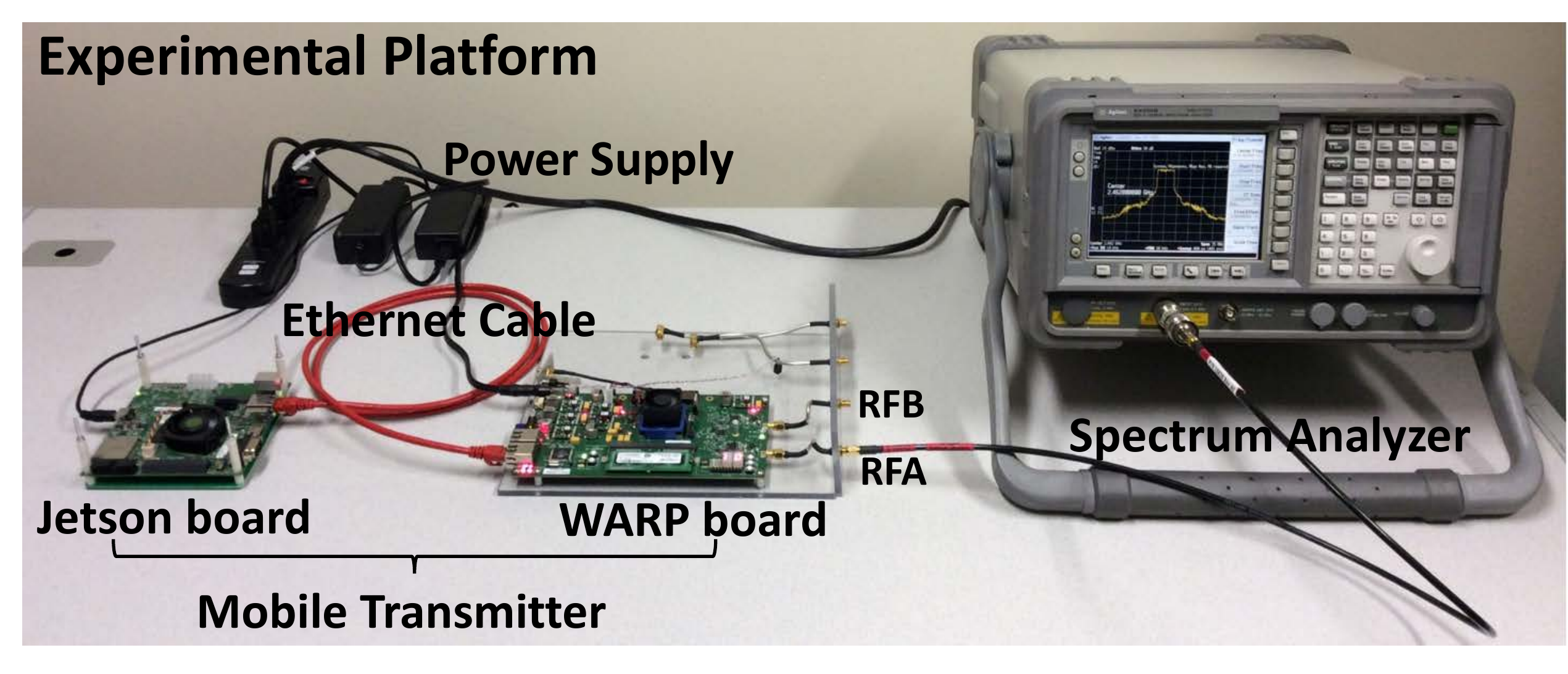}
\caption{Experimental setup}       % Give a unique label
\end{figure}

\section{Experimental Verification of DPD Functionality}
To verify the DPD suppression effect on spurious spectrum emissions experimentally, we stream the DPD output samples generated from Jetson board to WARP v3 radio board, which is equipped with MAX2829 transceiver and Anadigics AWL6951 PA, and monitor the PA output by an Agilent E4404B spectrum analyzer. 

Figure 5 shows the experimental setups, where we connect a Jetson board to WARP board via Ethernet cable as a software-defined mobile transmitter, and the radio output is sent to the spectrum analyzer via a coaxial cable. Here, the original transmit samples are generated on the Jetson board, and then passed through the DPD implementation accelerated on mobile GPU or multicore CPU, the DPD output samples will be packed into streaming packets, and sent to WARP based on socket APIs. The WARP board receives the packets from Jetson, processes those packets and finally directs them to the PA and RF by FPGA-based radio control modules. The underlying framework for packet transfer between Jetson and WARP and packet delivery to RF is based on WARPLab \cite{warplab}, with the original MATLAB-based baseband processing and socket wrapper replaced by CUDA-based or OpenMP/NEON-based DPD processing and a customized C-based socket wrapper, so that the input data can be processed and streamed to WARP at high data rate and monitored on spectrum analyzer in real time.

The DPD parameter estimation happens offline before we perform the actual DPD on streaming data as described above. For the offline training, we establish the feedback loop by connecting RF antenna connector A (RFA) and RF antenna connector B (RFB) on WARP, and collect the samples in the feedback path for estimation based on WARPLab.

The properties of PA on WARP can be formulated with a memoryless PA model, which is developed based on experimentally gathered PA input and output data:

\begin{align}
PA_{out}=\alpha_1 \cdot PA_{in}+\alpha_3|PA_{in}|^2PA_{in}+\alpha_5|PA_{in}|^4PA_{in},
\end{align}

Here, $\alpha_1=0.9490-0.0197i$, $\alpha_3=0.4885+0.1071i$, $\alpha_5=-1.0156-0.0474i$ are the 1$^{st}$, 3$^{rd}$, 5$^{th}$ polynomial coefficients for the 5-order WARP PA model.

\section{Performance Results}
\subsection{Data Rate Performance}
In this section, we benchmark the data rate performance of the DPD implementation on both mobile GPU and multicore CPU. The timing performance is measured by the CPU wall-clock latency $L$ for predistorting a certain number of input data. We need to ensure necessary synchronization between CPU and GPU by calling \texttt{cudaDeviceSynchronize()} when performing time measurement for the GPU implementation.

The computation workload scaled by the number of input samples $N$ and the clock frequency of the processor are two major factors which affect the data rate performance. Here, we calculate the throughput $T$ for the GPU implementation by:
\begin{align}
T_{GPU}= \frac{N\times (warpSize-R)}{L\times warpSize},
\end{align}
and for the CPU implementation simply by:
\begin{align}
T_{CPU}=\frac{N}{L},
\end{align}
where $R$ indicates the redundant threads in a $warp$ to facilitate the filtering computation, and is set to $R=L_p-1=L_q-1=4$ according to the design configuration in Table 2, and $warpSize=32$ is reserved by Nvidia.

\begin{figure*}[t]
\begin{subfigure}{.35\textwidth}
  %\centering
  \includegraphics[width=\linewidth]{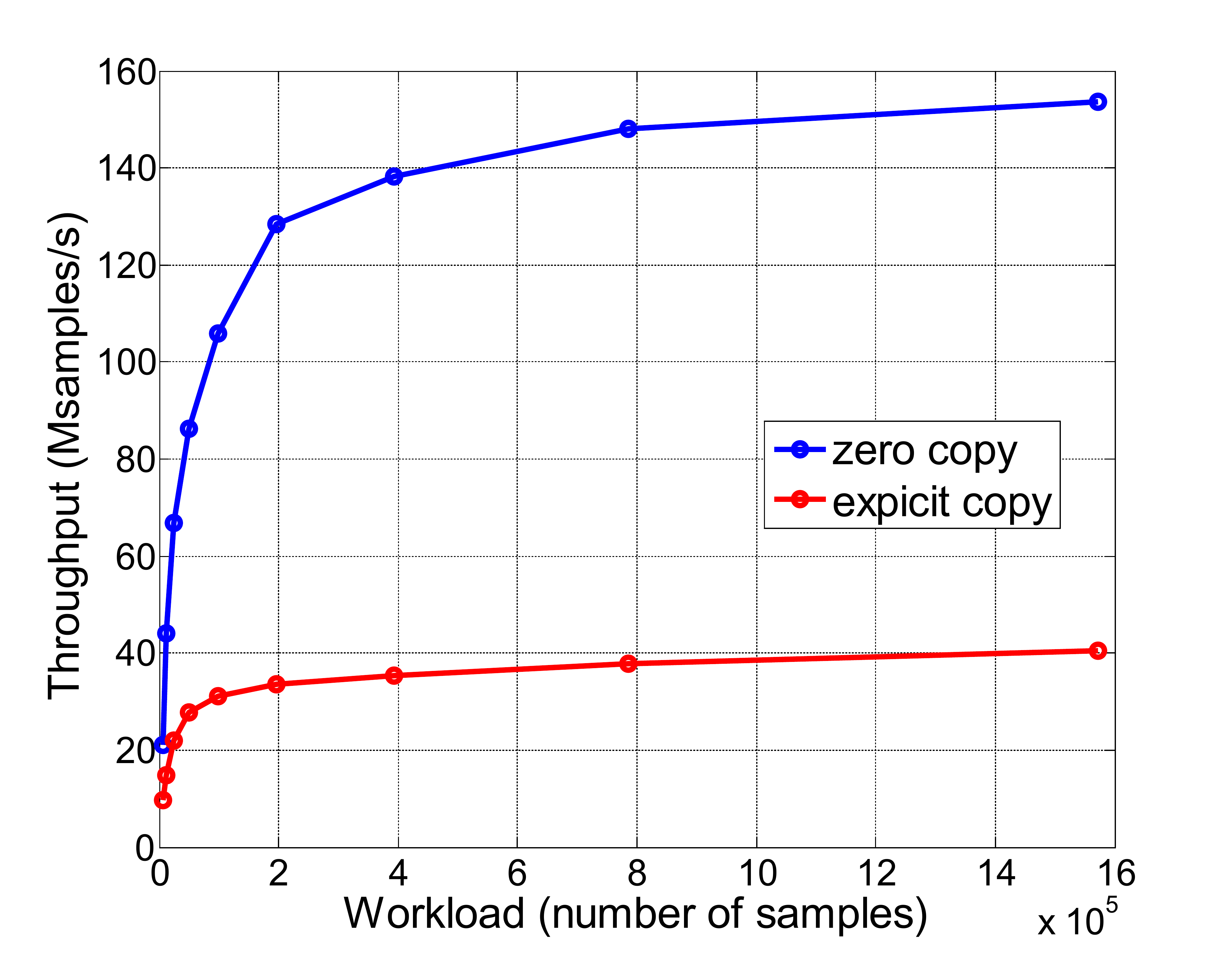}
  \caption{zero copy vs. explicit copy}
\end{subfigure}%
\begin{subfigure}{.35\textwidth}
  %\centering
  \includegraphics[width=\linewidth]{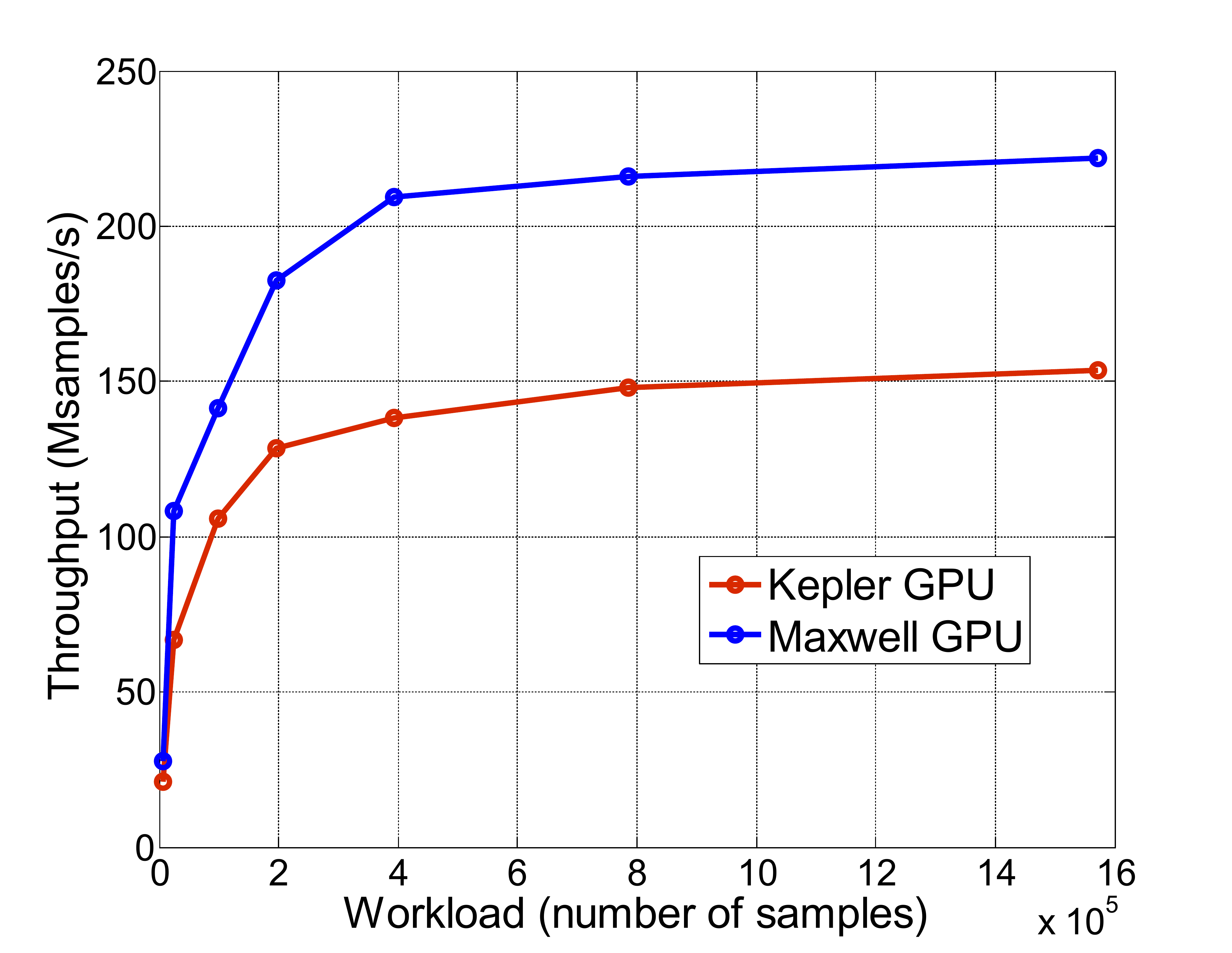}
  \caption{throughput vs. workload}
\end{subfigure}
\begin{subfigure}{.35\textwidth}
  %\centering
  \includegraphics[width=\linewidth]{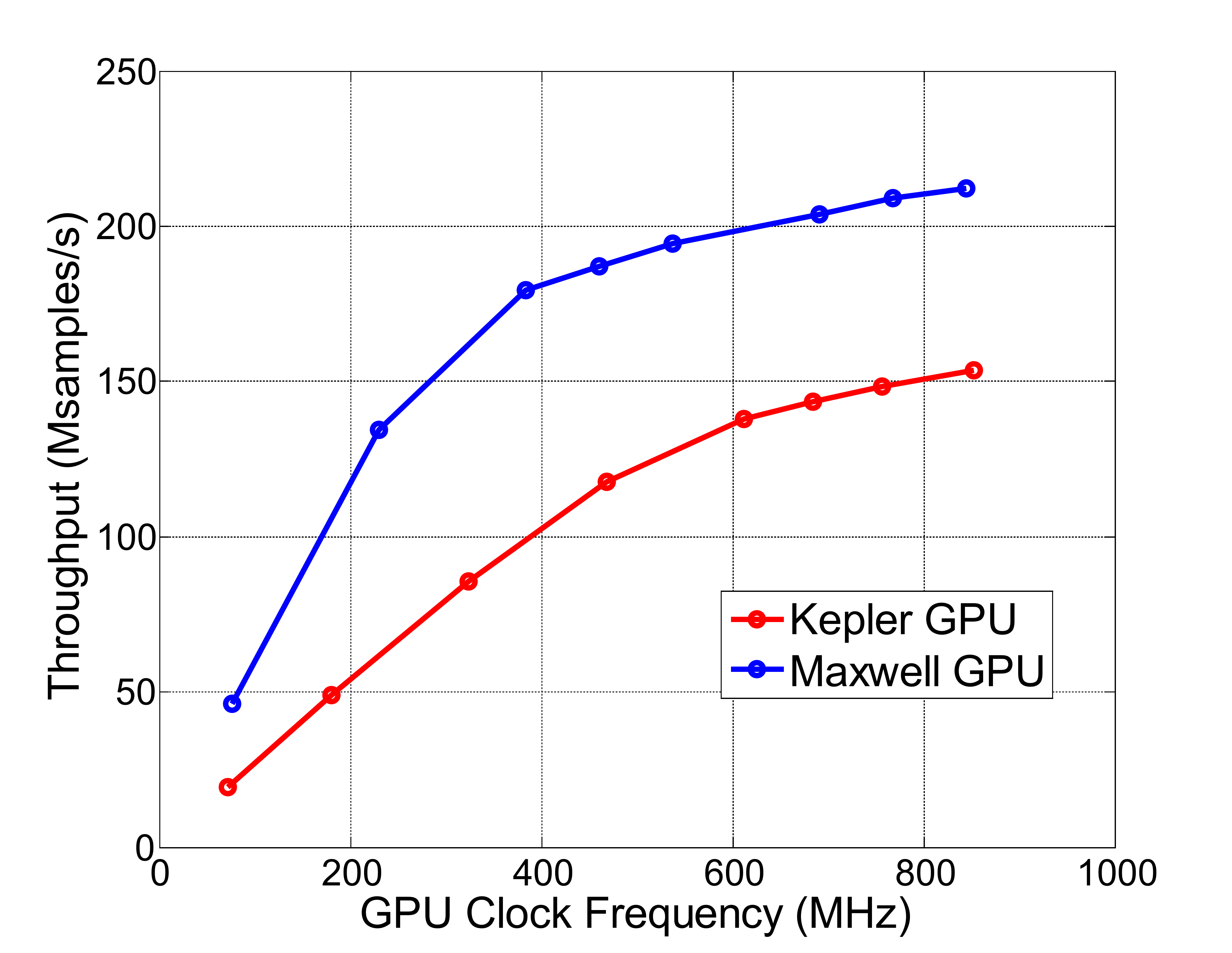}
  \caption{throughput vs. GPU freq.}
\end{subfigure}
\caption{Throughput performance of mobile GPU implementation}
\end{figure*}

\begin{figure*}[t]
\centering
\begin{subfigure}{.35\textwidth}
  %\centering
  \includegraphics[width=\linewidth]{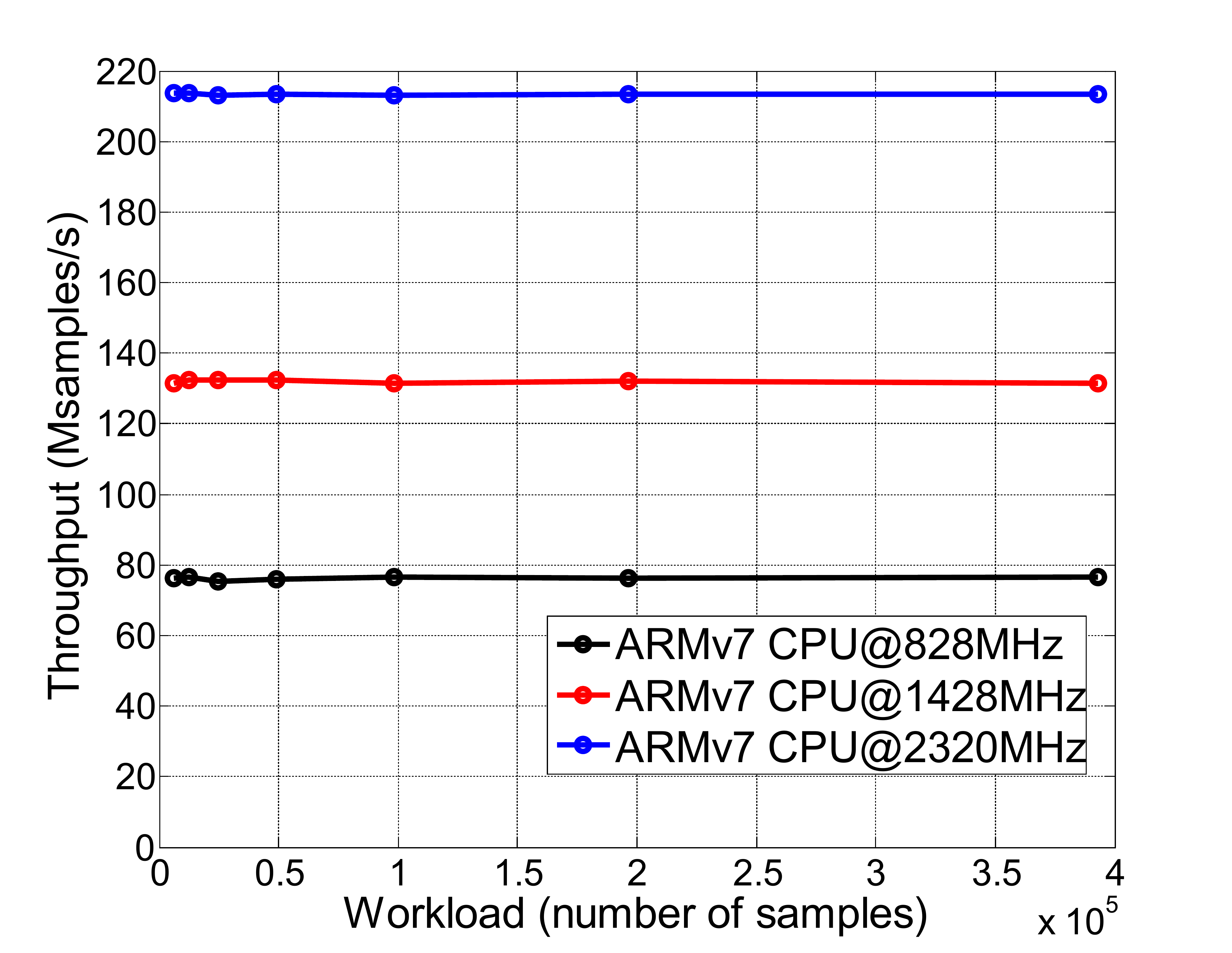}
  \caption{throughput vs. workload}
\end{subfigure}%
\begin{subfigure}{.35\textwidth}
  %\centering
  \includegraphics[width=\linewidth]{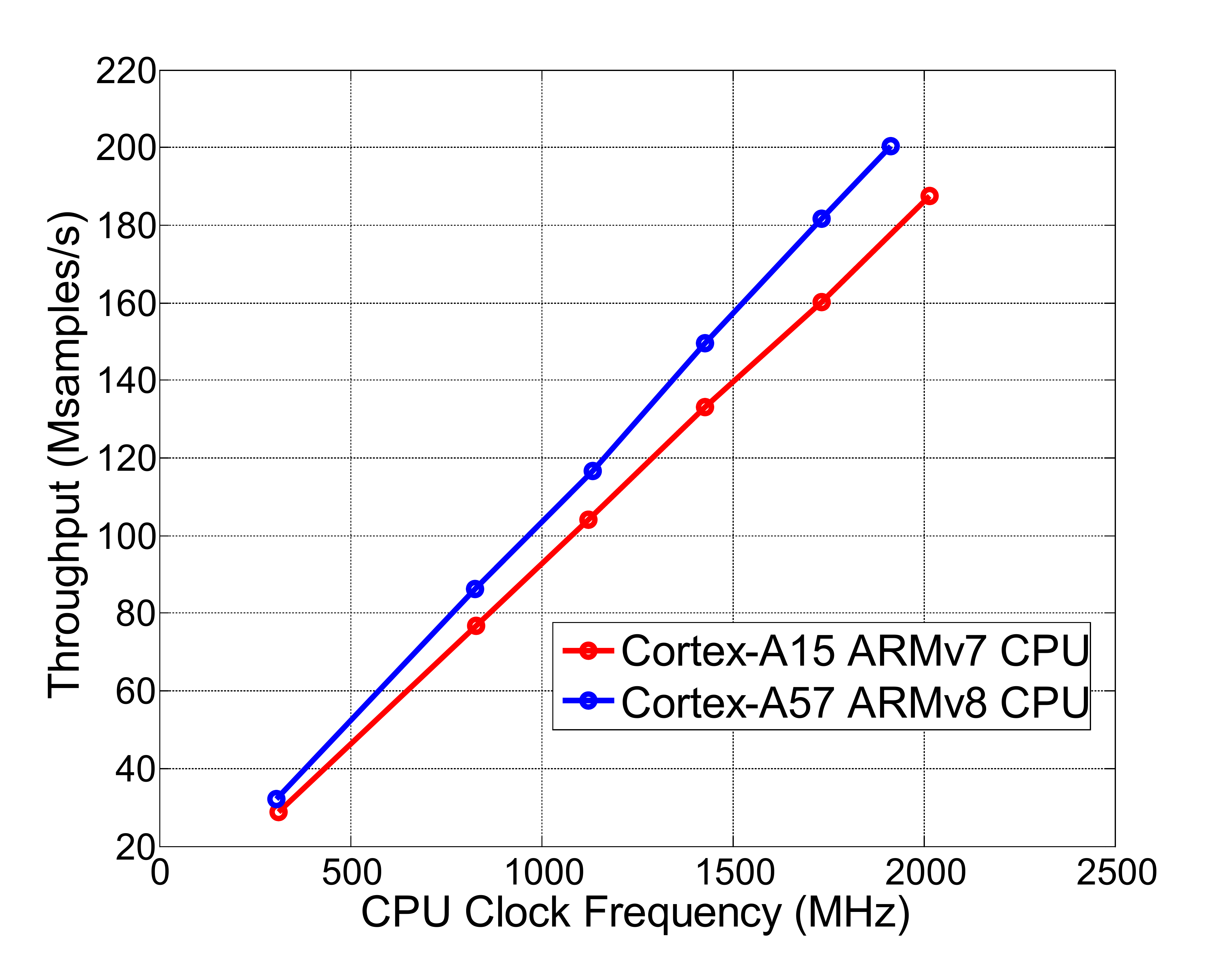}
  \caption{throughput vs. frequency}
\end{subfigure}
\caption{Throughput performance of mobile CPU implementation}
\end{figure*}

\begin{figure*}[t]
\centering
\begin{subfigure}{.45\textwidth}
  %\centering
  \includegraphics[width=\linewidth]{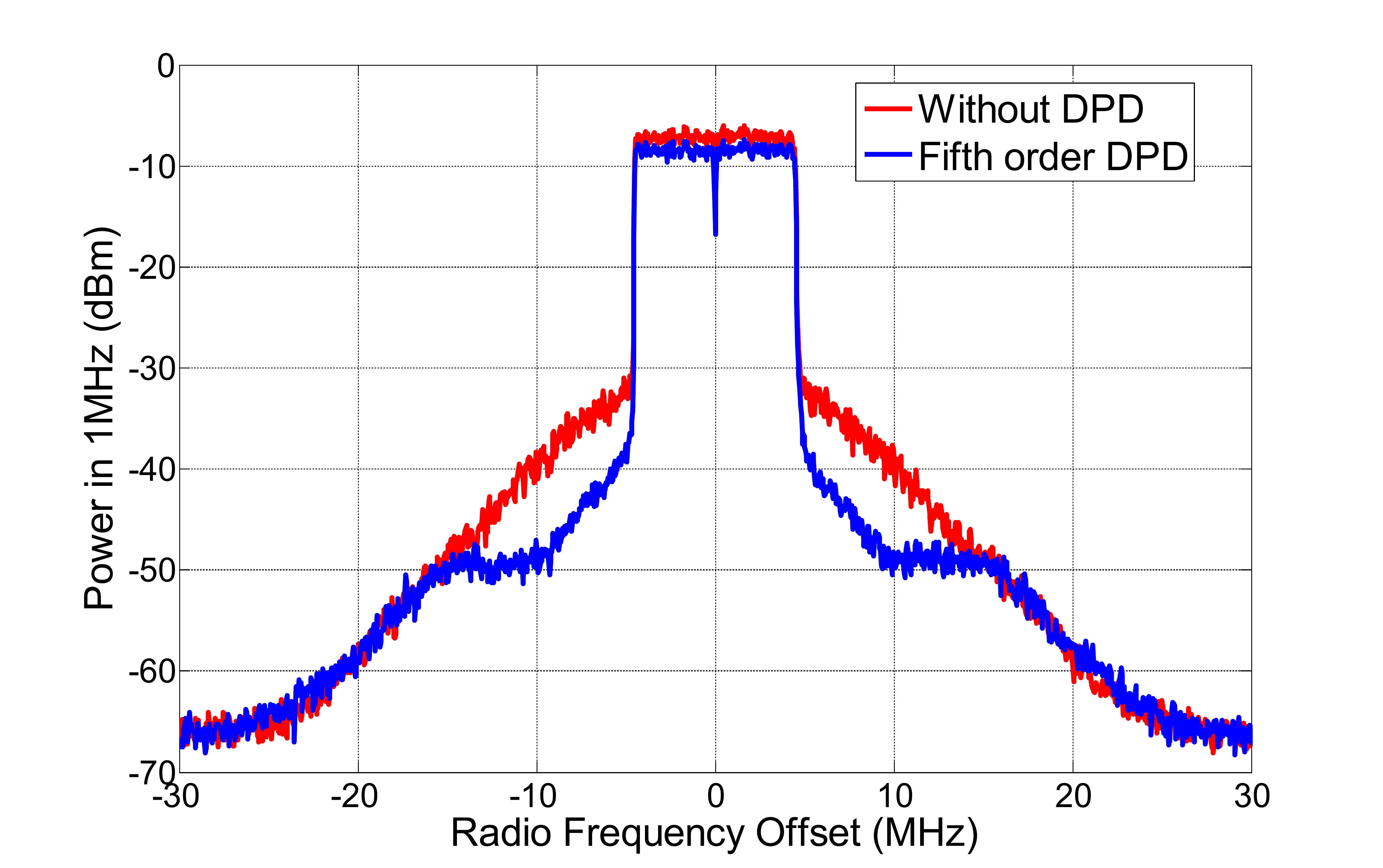}
  \caption{Single component carrier}
\end{subfigure}%
\begin{subfigure}{.45\textwidth}
  %\centering
  \includegraphics[width=\linewidth]{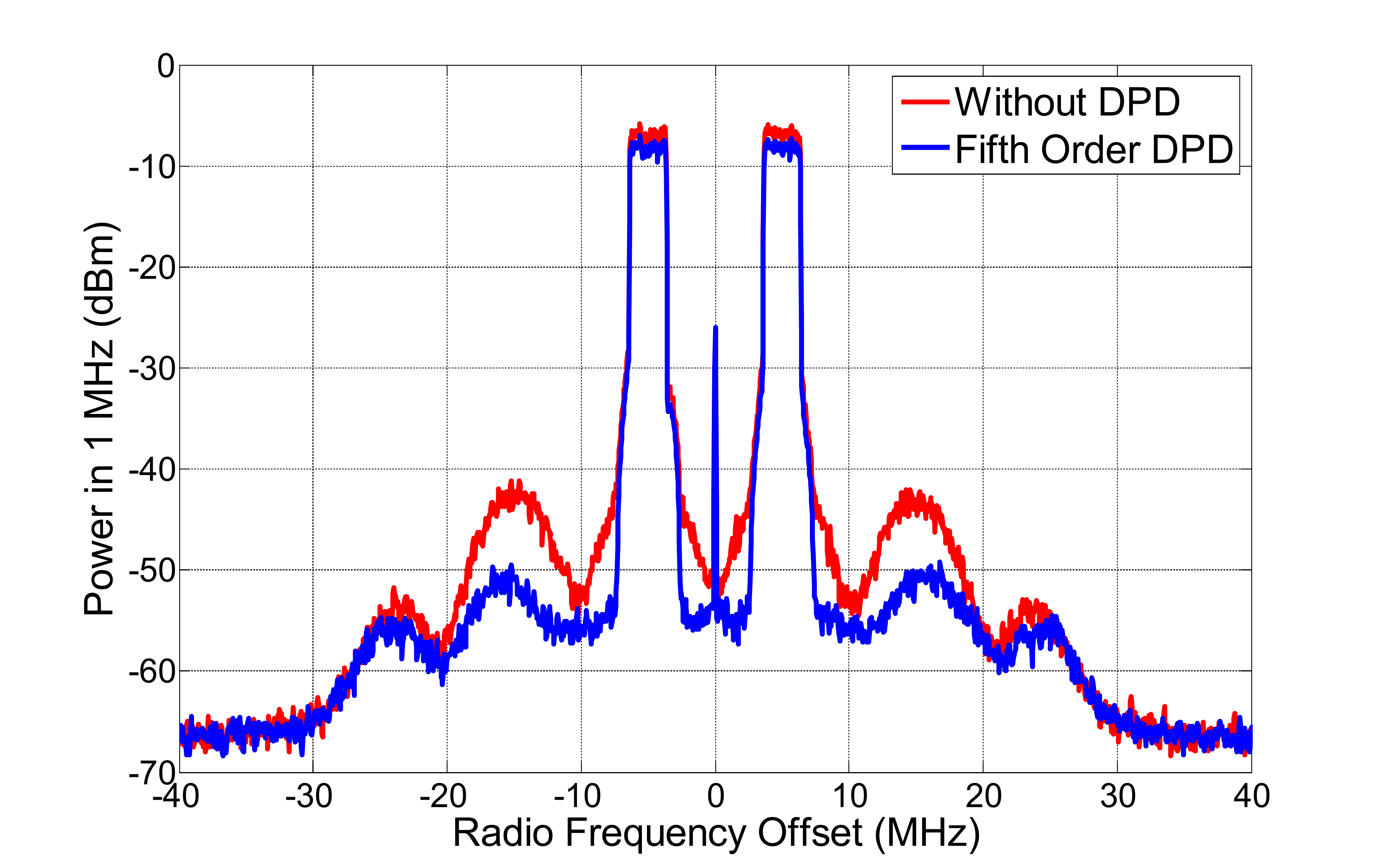}
  \caption{Non-contiguous carrier aggregation}
\end{subfigure}
\caption{DPD suppression effect}
\end{figure*}

In Figure 6(a), we show the throughput performance comparison between GPU implementations including explicit memory copy and zero copy, respectively. The benchmarks are performed with various workload $N$ on the Jetson TK1 board, and from the results we can find that explicit memory copy will poses significant overhead and performance degradation, while zero copy can achieve much higher performance with very low overhead.  

Figure 6(b) records the throughput performance comparison between the Kepler mobile GPU on Jetson TK1 board and the Maxwell mobile GPU on Jetson TX1 board, at various workload configurations with their maximum GPU clock frequencies, i.e, 852 MHz on TK1 and 998 MHz on TX1. With the increase of the workload $N$, the throughput performance increases on both Kepler GPU and Maxwell GPU, with increasing occupancy rates of computing cores, until saturation. Obviously, the Maxwell GPU can outperform the Kepler GPU because of more efficient streaming multiprocessors with higher maximum clock frequency, improved control logic partitioning, better workload balancing, and advanced instruction scheduling and issuing schemes. Our design achieves a peak performance over 150 Msamples/s on Jetson TK1 and over 220 Msamples/s on Jetson TX1.

Figure 6(c) records the throughput performance comparison between two boards at various GPU clock frequencies, which can be tuned manually \cite{freq}. We fix a large $N (N>1.5\times 10^6)$ to ensure high occupancy of GPU cores for all the frequency benchmarks. The throughput performance increases nearly linearly with the increase of frequency at low frequencies. At high frequencies, the throughput performance exposes  saturation due to thread deployment and scheduling overhead, and memory resource competitions and bandwidth limitations.

In Figure 7(a), we show the throughput performance comparison of the CPU implementation at different workloads. We verify various CPU frequencies on the Cortex-A15 CPU on the Jetson TK1 and conclude that the workload variation has little effect on the throughput performance on the CPU, since the four CPU cores can be easily saturated at low workloads.

Figure 7(b) shows the throughput performance comparison between two boards at various CPU clock frequencies. Here, we fixed a large workload, for example, $N=1\times 10^5$, for more consistent and stable results. We find that the Cortex-A57 CPU with ARMv8 architecture on TX1 achieves higher throughput performance than Cortex-A15 CPU with ARMv7 architecture on TK1 at similar frequencies, and the throughput can increase almost linearly with the CPU frequency, to a peak performance of around 200 Msamples/s.

In Table 4, we compare the peak performances of our implementations with previous work. Our work exceeds previous work in terms of throughput performance and sample precision.

Besides data rate performance, energy efficiency is another dimension of design space exploration, especially for mobile devices where power is usually constrained. To enable high energy efficiency, Tegra K1 and Tegra X1 SoCs are designed for the mobile market with a peak power consumption at the level of 10W-15W\cite{tk1,tx1}. Some previous work has presented power measurements of Tegra K1 and X1 using power meters\cite{power_measure}, which indicates that mobile GPUs usually enable higher performance/watt than embedded CPUs because of high throughput with low clock frequencies. Some other work has discussed accurate power modeling for Tegra SoCs\cite{power_model}, showing that the power consumption of Tegra SoCs can be higher with the increase of off-chip memory clock frequencies, GPU/CPU hardware utilization such as on-chip local caches and functional units, as well as GPU/CPU clock frequencies. To arrive at optimal performance/watt of our DPD designs, we need to further perform detailed benchmarks on hardware utilization at each single step of computation with experimental measurements of the power consumption, which can be an interesting following work in the future.

\begin{table}

\centering
\caption{Performance comparison with previous work}

\scriptsize

\begin{tabular}{c|c|c}
\hline 
 & \textbf{Data precision} & \textbf{Peak throughput}\tabularnewline
\hline 
\hline 
90nm CMOS TTA ~\cite{tta} & 16-bit FP & 20.8 Msamples/s\tabularnewline
\hline 
45nm CMOS TTA ~\cite{tta} & 16-bit FP & 53.3 Msamples/s\tabularnewline
\hline 
Kepler GPU @ 852MHz & 32-bit FP & 153.5 Msamples/s\tabularnewline
\hline 
Maxwell GPU @ 998MHz & 32-bit FP & 221.8 Msamples/s\tabularnewline
\hline 
ARMv7 CPU @ 2.32GHz & 32-bit FP & 214.4 Msamples/s\tabularnewline
\hline 
ARMv8 CPU @ 1.91GHz & 32-bit FP & 200.1 Msamples/s\tabularnewline
\hline 
\end{tabular}
\end{table}

\subsection{DPD Suppression Performance}
On the mobile transmitter built by Jetson board and WARP radio, we experimentally record the PA output to verify the DPD suppression effect on spurious spectrum emissions. We prepare a single 10MHz LTE uplink carrier, as well as two non-contiguous 3MHz LTE uplink carriers with 10MHz spacing as original input of DPD, to show the DPD effect for single-carrier and non-contiguous carrier aggregation scenarios. Figure 8(a) and Figure 8(b) show that we can experimentally achieve over 10dB suppression on major spurious spectrum components, indicating that our DPD design works effectively for practical radio setups.

\section{Conclusion}
In this paper, we present high performance parallel DPD implementations on mobile GPU and embedded multicore CPU for mobile transmitters. For both GPU and CPU implementations, we explore and realize the inherent data parallelism of DPD based on corresponding parallel programming models and schemes. To reduce the data sharing and communication overhead in our implementations, we take advantage of \emph{warp shuffle} techniques on GPU and vector extraction intrinsic on CPU to resolve the data dependencies efficiently within local registers. Our DPD implementations achieve over 150 Msamples/s peak throughput on a Kepler mobile GPU, over 220 Msamples/peak throughput on a Maxwell mobile GPU, and over 200 Msamples/s peak throughput on ARMv7 and ARMv8 multicore CPU, which indicates that our DPD design can efficiently support practical high bandwidth mobile transmitters. By integrating our DPD implementation on a customized software-defined mobile transmitter built by Jetson board and WARP radio board, we further experimentally verify that our DPD design can suppress major spurious spectrum emissions effectively by over 10dB on real radio hardware.

\begin{acknowledgements}
This work was supported by the US NSF under grants EECS-1408370, CNS-1265332, ECCS-1232274, and the Finnish Agency of Innovation, Tekes.
\end{acknowledgements}

\bibliographystyle{IEEEbib} 
\bibliography{dpd}   % name your BibTeX data base

\iffalse

\fi
\end{document}